\newcommand{\version}{June 3, 2003}
\documentclass[12pt]{article}
\usepackage{amsthm,amsfonts,amssymb,latexsym,amsmath}

{\catcode `\@=11 \global\let\AddToReset=\@addtoreset}
\AddToReset{equation}{section}

\theoremstyle{plain}
\newtheorem{thm}{THEOREM}
\newtheorem{cor}{COROLLARY}
\newtheorem{lem}{LEMMA}
\newtheorem{prop}{PROPOSITION}
\theoremstyle{definition}

\newcommand{\beq}{\begin{equation}}
\newcommand{\eeq}{\end{equation}}
\def\beqa{\begin{eqnarray}}
\def\eeqa{\end{eqnarray}}
\newcommand{\infspec}{{\rm inf\ spec\ }}
\newcommand{\R}{{\mathbb R}}

\newcommand{\N}{{\mathbb N}}
\newcommand{\Z}{{\mathbb Z}}

\newcommand{\Hh}{{\mathcal H}}

\newcommand{\eps}{\varepsilon}
\newcommand{\xx}{\mathord{\hbox{\boldmath {\scriptsize{$x$}}}}}
\newcommand{\x}{\mathord{\hbox{\boldmath $x$}}}
\newcommand{\y}{\mathord{\hbox{\boldmath $y$}}}
\newcommand{\yy}{\mathord{\hbox{\boldmath {\scriptsize{$y$}}}}}
\newcommand{\X}{\mathord{\hbox{\boldmath $X$}}}
\newcommand{\0}{\mathord{\hbox{\boldmath $0$}}}
\newcommand{\Tr}{{\rm Tr}}
\newcommand{\half}{\mbox{$\frac{1}{2}$}}
\newcommand{\E}{{\mathcal E}^{\rm GP}}
\newcommand{\F}{{\mathcal F}}

\newcommand{\Engp}{E^{\rm GP}}
\newcommand{\Endm}{E^{\rm DM}}
\newcommand{\pgp}{\phi^{\rm GP}}
\newcommand{\Enmod}{\widetilde E^{\rm DM}}
\newcommand{\Edm}{{\mathcal E}^{\rm DM}}
\newcommand{\Edmmod}{\widetilde{\mathcal E}^{\rm DM}}
\newcommand{\rdm}{\rho^{\rm DM}}
\newcommand{\gdm}{\gamma^{\rm DM}}
\newcommand{\mdm}{\mu^{\rm DM}}

\newcommand{\Pdm}{P^{\rm DM}}
\newcommand{\Eqm}{E^{\rm QM}}
\newcommand{\bos}{{\rm bose}}
\newcommand{\bfom}{\mathord{\hbox{\boldmath $\Omega$}}}
\newcommand{\bfL}{\mathord{\hbox{\boldmath $L$}}}
\newcommand{\bfnab}{\mathord{\hbox{\boldmath $\nabla$}}}
\newcommand{\bfA}{\mathord{\hbox{\boldmath $A$}}}
\newcommand{\bfomm}{\mathord{\hbox{\boldmath {\scriptsize{$\Omega$}}}}}
\newcommand{\ham}{H_{N,\bfomm,a}}
\newcommand{\Hmod}{\widetilde H}
\newcommand{\rmod}{\widetilde\rho}
\newcommand{\gmod}{\widetilde\gamma}
\newcommand{\Pmod}{\widetilde P}
\newcommand{\matB}{\mathord{\hbox{\boldmath $B$}}}
\newcommand{\matell}{\mathord{\hbox{\boldmath $\ell$}}}
\newcommand{\matrho}{\mathord{\hbox{\boldmath $\rho$}}}
\newcommand{\matW}{\mathord{\hbox{\boldmath $W$}}}
\newcommand{\matn}{\mathord{\hbox{\boldmath $n$}}}
\newcommand{\const}{{\rm const. \, }}
\newcommand{\NN}{{\cal N}}
\newcommand{\Oset}{{\cal O}}

\date{\small\version}
\begin{document}
\markboth{\scriptsize{R. Seiringer, Rotating Gas,
\version}}{\scriptsize{R. Seiringer, Rotating Gas, \version}}
\title{{\bf Ground State Asymptotics of a Dilute, Rotating Gas}}

\author{\vspace{5pt} Robert Seiringer
\\ \vspace{-4pt}\small{ Department of Physics, Jadwin Hall, Princeton
University}\\ \vspace{-4pt}\small P.O. Box 708, Princeton, NJ
08544, USA
\\ {\small Email: \texttt  {rseiring@math.princeton.edu}} }
\maketitle 

\begin{abstract}
We investigate the ground state properties of a gas of interacting
particles confined in an external potential in three dimensions and
subject to rotation around an axis of symmetry. We consider the
so-called Gross-Pitaevskii (GP) limit of a dilute gas. Analyzing both
the absolute and the bosonic ground state of the system we show, in
particular, their different behavior for a certain range of
parameters. This parameter range is determined by the question whether
the rotational symmetry in the minimizer of the GP functional is
broken or not. For the absolute ground state, we prove that in the GP
limit a modified GP functional depending on density matrices correctly
describes the energy and reduced density matrices, independent of
symmetry breaking. For the bosonic ground state this holds true if and
only if the symmetry is unbroken.
\end{abstract}

\section{Introduction}

Since the first experimental realization of Bose-Einstein condensation
(BEC) in dilute gases of alkali atoms \cite{cornell,ketterle}, much
interest has been devoted to the study of their rotational
properties. Beautiful results showing the appearance of vortices and
the formation of vortex arrays have been obtained in various
experiments \cite{MCWD00,ARVK01}. Many of the striking features of
rotating Bose-Einstein condensates are well described by means of the
Gross-Pitaevskii (GP) functional, and most theoretical investigations
rely on the approximation made in its use (see, e.g.,
\cite{FS01,dalfovo}).

Recently a rigorous justification of the GP functional for
non-rotating systems has been obtained \cite{LSY00}, and a proof of
BEC in the ground state was given in \cite{LS02}. These results do not
extend to the case of a rotating system in a simple way, however. In
this paper, we investigate the effect of the rotation. Several new
features come into play. First, Bose statistics become
essential. While for the case of a non-rotating system the ground
state of the Hamiltonian is automatically symmetric in the particle
coordinates, this in not necessarily the case for rotating systems
and, in fact, is shown not to be the case for the system under
consideration here (at least for a certain parameter range). That is, it
is important to distinguish the absolute ground state from the bosonic
ground state (the ground state of the Hamiltonian restricted to totally
symmetric wave functions), the two having significantly different physical
properties. Secondly, the appearance of vortices breaks the rotational
symmetry, leading to non-uniqueness of the minimizer of the GP
functional (see Theorem~\ref{T3} below). This makes it necessary to
study a generalized GP functional, which depends on density matrices
rather than densities alone. These two properties are in fact related,
as we show, the symmetry being broken if and only if the absolute and
bosonic ground state energy differ by a significant amount (of the
order of the energy itself).

The main result of this paper is that the generalized GP density
matrix functional mentioned above correctly describes the absolute
ground state of the Hamiltonian under consideration here, in a certain
dilute limit. More precisely, it gives the correct asymptotics of the
ground state energy and all the $n$-particle reduced density matrices
of the ground state of the system. In the case of unbroken rotational
symmetry, the GP density matrix functional agrees with the usual GP
functional, and these results apply also to the bosonic ground
state. This extends previous results on superfluidity in
\cite{LSYsfl}. In the general case of broken symmetry, however, we are
not able to establish the precise asymptotics in the bosonic case, but
we give bounds on the corresponding energy that show in particular
that it's energy differs from the absolute ground state energy by a
significant amount. Moreover, this implies that the absolute ground
state has a huge degeneracy that increases exponentially with the
particle number.

Before we can give the precise formulation of the preceding
statements, we have to define the system under consideration. For
$\bfom\in\R^3$ and a realvalued $V\in L^\infty_{\rm loc}(\R^3,d\x)$
let $H_0$ be the one-particle Hamiltonian
\beq
H_0=-\Delta-\bfom\cdot \bfL + V(\x) \ ,
\eeq
acting on $L^2(\R^3,d\x)$. Here $\Delta=\bfnab^2$ is the Laplacian on
$\R^3$, and $\bfL=-i \x\wedge \bfnab$ denotes the angular momentum
operator. The operator $H_0$ is the appropriate Hamiltonian for
one particle confined in a trap potential $V$ and rotating with
angular velocity $\bfom$, in the rotating frame. To ensure that $H_0$
is semibounded from below, we assume that
\beq\label{12}
\Big[ V(\x)- \frac{\omega^2}4 r^2\Big]_- \in L^\infty(\R^3)
\eeq
for all $\omega$ in some non-zero interval $[0,\Omega_c)\subset
\R_+$, where we denote by $[\, \cdot \,]_-$ the negative part, and
$r$ denotes the radial distance perpendicular to $\bfom$, i.e.,
$r=|\x\wedge\bfom|/|\bfom|$. We call the largest possible
$\Omega_c$ the {\it critical angular velocity}, allowing it to be
infinity. Writing $H_0$ as
\beq
H_0= (-i \bfnab - \bfA )^2 + V(\x) - \frac {|\bfom|^2}{4} r^2 \ ,
\eeq
where $\bfA= \half \bfom\wedge\x$ is the vector potential of a
constant magnetic field in the direction of $\bfom$, we see that $H_0$
is semibounded from below for $|\bfom|<\Omega_c$, and unbounded for
$|\bfom|>\Omega_c$. We will henceforth always assume that
$|\bfom|<\Omega_c$. Note that the condition (\ref{12}) on $V$, as well
as $\Omega_c$, depend on the direction of~$\bfom$, which we assume to
be fixed throughout the paper, whereas its absolute value is allowed
to vary.

We also assume that $\lim_{|\xx|\to\infty} V(\x)=\infty$ uniformly
in all directions, implying discrete spectrum of $H_0$. Moreover,
we demand that $V$ commutes with $\bfom\cdot \bfL$, or, in other
words, it is {\it axially symmetric}, being a function that
depends only on $r$ and $z=\x\cdot\bfom /|\bfom|$. Without loss of
generality $V\geq 0$.

To define the $N$-particle problem, consider the Hilbert space
$\Hh_N=\bigotimes_{i=1}^N L^2(\R^3,d\x_i)$, the $N$-fold tensor
product of the one-particle space. The Hamiltonian corresponding
to $N$ particles in a trap $V$, rotating with angular velocity
$\bfom$, and interacting pairwise with an interaction potential
$v$, is given by
\beq\label{ham}
H_{N,\bfomm,a}=\sum_{i=1}^N H_0^{(i)} + \sum_{1\leq i<j\leq N }
\frac{1}{a^2} v\big( (\x_i-\x_j)/a\big) \ .
\eeq
Here the superscript $(i)$ means that $H_0$ acts on the $i$'th
part in the tensor product. The interaction potential $v$ is
assumed to positive, spherically symmetric and of compact support.
We do not demand it to be integrable, it is allowed to have a hard
core, which reduces the domain of definition of $\ham$ to wave
functions in $\Hh_N$ that vanish whenever two particles are closer
together then the size of the hard core. The positive parameter
$a$ appearing in $\ham$ determines the range of the interaction
$v$. We assume that $v$ has {\it scattering length} $1$ (see
\cite{LSY00} or \cite{LY01} for a definition), implying that
$a^{-2} v(\x/a)$ has scattering length $a$.

We are interested in the ground state properties of $\ham$ for large
$N$ and small $a$. In particular, the {\it Gross-Pitaevskii limit}
$a\sim N^{-1}$ will be investigated. Its significance comes from the
fact that it ensures that the contributions to the energy of all the
terms in the Hamiltonian (\ref{ham}) are of the same order as $N\to
\infty$.  Let $\Eqm$ denote the ground state energy of $\ham$, i.e.,
\beq\label{eqm}
\Eqm(N,\bfom,a)=\infspec H_{N,\bfomm,a} \ .
\eeq
Besides $N$, $\bfom$ and $a$ it depends on $V$ and $v$, of course, but
these potentials are assumed to be fixed once and for all. From the
discussion above it is clear that $\Eqm$ is finite for
$|\bfom|<\Omega_c$, and that $\Eqm(N,\bfom,a)=\Eqm(N,-\bfom,a)$.

A different problem is obtained by restricting ourselves to wave
functions in $\Hh_N$ that are totally symmetric with respect to
exchange of two particle coordinates, which corresponds to
assuming the particles to be Bosons. We denote the corresponding
ground state energy by $\Eqm_\bos(N,\bfom,a)$, i.e., 
\beq\label{eqmbos}
\Eqm_\bos(N,\bfom,a)=\inf \Big\{ \langle\Psi |\ham\Psi\rangle \, :
\, \Psi\in P_\bos\Hh_N, \, \|\Psi\|_2=1 \Big\} \ ,
\eeq
where $P_\bos$ denotes the projection onto totally symmetric
functions. The two quantities (\ref{eqm}) and (\ref{eqmbos}) a
{\it a priori} unrelated, except for the trivial inequality
$\Eqm(N,\bfom,a)\leq \Eqm_\bos(N,\bfom,a)$. For $\bfom=\0$, i.e.,
the case of a Schr\"odinger operator, it is well known that
$\Eqm(N,\0,a)=\Eqm_\bos(N,\0,a)$, but for $\bfom\neq \0$ this need
not necessarily be the case. In fact it will turn out that, at
least for certain values of the parameters, the two quantities are
different.

Before we can state our main results, we have to introduce some
functionals which will turn out to be related to the
asymptotic behavior of the $N$-particle problem defined by $\ham$
for large $N$ and small $a$. We do this in the next section.

\section{Gross-Pitaevskii Functionals}

We define the {\it Gross-Pitaevskii density matrix} (DM) functional to
be
\beq\label{defdm}
\Edm_{\bfomm,g}[\gamma]=\Tr [ H_0\gamma] + 4\pi  g \int
\rho_\gamma(\x)^2 d\x \ .
\eeq
Here $\gamma$ is a one-particle density matrix, a positive
trace-class operator on $L^2(\R^3,d\x)$, and $\rho_\gamma$ denotes
its density. The corresponding ground state energy, the infimum of
(\ref{defdm}) under the condition $\Tr[\gamma]=1$, will be denoted
by $\Endm(\bfom,g)$, i.e.,
\beq\label{endm}
\Endm(\bfom,g)= \inf \Big\{ \Edm_{\bfomm,g}[\gamma] \, : \,
\Tr[\gamma]=1 \Big\} \ .
\eeq
Note that one could equivalently define the ground state energy
under the subsidiary condition $\Tr[\gamma]=N$, or any other
constant, but this can be easily related to (\ref{endm}) by a
trivial scaling of $\Endm$ and $g$. Again it is clear that
$\Endm(\bfom,g)$ is finite for $|\bfom|<\Omega_c$ and $g\geq 0$,
which we will always assume.

The analogue of the DM functional for a two-dimensional gas was
introduced in \cite{S02} as a generalization of the standard
GP functional, which is given by restricting
$\Edm_{\bfomm,g}$ to density matrices of rank one. Equivalently,
one can write
\beq\label{gpfunct}
\E_{\bfomm,g}[\phi]=\langle \phi| H_0\phi\rangle + 4\pi g
\int|\phi(\x)|^4 d\x
\eeq
for functions $\phi\in L^2(\R^3,d\x)$, and define the
corresponding ground state energy as
\beq\label{engp}
\Engp(\bfom,g)= \inf \Big\{ \E_{\bfomm,g}[\phi] \, : \,
\|\phi\|_2=1 \Big\} \ .
\eeq
Since one-dimensional projections are legitimate density matrices in
(\ref{defdm}), it is clear that $\Endm(\bfom,g)\leq \Engp(\bfom,g)$.

In the case of two dimensions both the functionals $\Edm_{\bfomm,g}$
and $\E_{\bfomm,g}$ were studied in \cite{S02}. Many of the results
translate directly to the three-dimensional case, with minor
modifications, and we merely state them here in the following two
propositions, omitting the proofs.

\begin{prop}[Minimizer of 
${\mathord{\hbox{\boldmath $\Edm_{\bfomm,g}$}}}$]\label{T1}
For each $0\leq |\bfom|<\Omega_c$ and $g> 0$ there exists a unique
minimizing density matrix for (\ref{defdm}) under the condition
$\Tr[\gamma]=1$, denoted by $\gdm_{\bfomm,g}$. This minimizer also
minimizes the linearized functional
\beq\label{deflin}
\gamma\mapsto\Tr [ (H_0+8\pi g \rdm_{\bfomm,g})\gamma] 
\eeq
(under the same normalization condition), where $\rdm_{\bfomm,g}$
denotes the density of $\gdm_{\bfomm,g}$.  Moreover, $\gdm_{\bfomm,g}$
has finite rank. Its density $\rdm_{\bfomm,g}$ is a bounded function, with
$\rdm_{\bfomm,g}(\x)\leq
\mdm_{\bfomm, g}/(8\pi g)$, where $\mdm_{\bfomm,g}$ is the
chemical potential in DM theory, which is the ground state energy
of (\ref{deflin}).
\end{prop}

Note that for uniqueness $g>0$ is essential, since the ground state of
$H_0$ may be degenerate, implying non-uniqueness for $g=0$.

Because of uniqueness, we also know that $\rdm_{\bfomm,g}$ is
axially symmetric, with symmetry axis $\bfom$. Note that the
uniqueness of $\gdm_{\bfomm,g}$ is non-trivial. From the strict
convexity of $\Edm_{\bfomm,g}$ in $\rho_\gamma$ it follows only
that $\rdm_{\bfomm,g}$ is unique, i.e., that every possible
minimizer has the same density. By examining the ground state
space of the operator appearing in (\ref{deflin}) one then shows
that there can be only one density matrix with the property
that it minimizes (\ref{deflin}) {\it and} has $\rdm_{\bfomm,g}$
as its density.

For $\E_{\bfomm,g}$ essentially the same results as in
Prop.~\ref{T1} are true, except for uniqueness.

\begin{prop}[Minimizers of 
${\mathord{\hbox{\boldmath $\E_{\bfomm,g}$}}}$]\label{T2}
For each $0\leq |\bfom|<\Omega_c$ and $g\geq 0$ there exists a
minimizing function for (\ref{gpfunct}) under the condition
$\|\phi\|_2=1$. Any minimizer $\phi$ fulfills the GP equation
\beq\label{gpeq}
H_0 \phi+8\pi g|\phi|^2\phi = \mu^{\rm GP}_{\bfomm,g} \phi \ ,
\end{equation}
where $\mu^{\rm GP}_{\bfomm,g}=\mu^{\rm GP}_{\bfomm,g}(\phi)$ is
the chemical potential, given by
\begin{equation}\label{mugp}
\mu^{\rm GP}_{\bfomm,g}=E^{\rm GP}(\bfom,g)+ 4\pi g\int
|\phi(\x)|^4 d\x \ .
\end{equation}
\end{prop}

One might suspect that the two minimization problems (\ref{endm}) and
(\ref{engp}) are equivalent, in the sense that the minimizer of
$\Edm_{\bfomm,g}$ has rank one, i.e, $\gdm_{\bfomm,g}$ is a
one-dimensional projection onto the minimizer of $\E_{\bfomm,g}$,
which would consequently have to be unique (up to a constant phase
factor, of course). However, for $\bfom\neq \0$ this is not true, at
least not for $g$ large enough. The following Theorem~\ref{T3} is the
analogue of Theorem~4 and Corollary~2 in \cite{S02}, where the case of
a two-dimensional system was considered. Since its proof is not a
simple generalization of the two-dimensional case, it will be given in
the Appendix. Due to some technical complications it is more difficult
to prove this theorem in three dimensions, and it will be convenient
not to consider the most general external potentials $V$, but to
restrict ourselves to a special class with sufficiently nice
properties. The general class of $V$'s which our proof applies to is
rather difficult to characterize, but sufficient conditions are easy
to state. We assume them for simplicity, they are general enough to
allow for a quite large class of $V$'s, but they are by no means
necessary for Theorem~\ref{T3} to hold, as the proof in the appendix
shows.

For Theorem~\ref{T3} (and only there) we will assume that the
external potential $V$ fulfills the bounds
\beq\label{b1}
V(\x)\leq \const (1+ r^s + |z|^p)
\eeq
and
\beq\label{b2}
V(\x) \geq \const (r^s + |z|^p) - \const
\eeq
for suitable constants (independent of $r$ and $z$) and for some
exponents $2\leq s<\infty$ and $0<p<\infty$. Note that $s\geq 2$ is necessary for condition (\ref{12}) to hold. 

\begin{thm}[Non-equivalence of 
${\mathord{\hbox{\boldmath $\E_{\bfomm,g}$}}}$ and 
${\mathord{\hbox{\boldmath $\Edm_{\bfomm,g}$}}}$]
\label{T3}
Assume that $V$ satisfies the bounds (\ref{b1}) and (\ref{b2}).  For
any $0<|\bfom|< \Omega_c$ there exists a $g_{|\bfomm|}$ such that
$g\geq g_{|\bfomm|}$ implies that no minimizer of $\E_{\bfomm,g}$ is an
eigenfunction of the angular momentum $\bfom\cdot \bfL$, and
consequently the minimizer is not unique. Moreover, for $g\geq g_{|\bfomm|}$,
$\Engp(\bfom,g)>\Endm(\bfom,g)$, and the minimizer $\gdm_{\bfomm,g}$
of $\Edm_{\bfomm,g}$ has at least rank 2.
\end{thm}

For a given direction of $\bfom$, we will denote by $\Xi\subset
[0,\Omega_c)\times \R_+ $ the set of parameters where $\E_{\bfomm,g}$
and $\Edm_{\bfomm,g}$ are not equivalent, i.e.,
\beq\label{defxi}
\Xi \equiv \left\{ (|\bfom|,g) \, : \, 
\Engp(\bfom,g)>\Endm(\bfom,g)\right\}\ .
\eeq
This is the case if and only if the rank of $\gdm_{\bfomm,g}$ is
greater or equal to two, and therefore, by Prop.~\ref{T1}, the ground
state of $H_0+8\pi g\rdm_{\bfomm,g}$ is degenerate. Note that
Theorem~\ref{T3} states that $\Xi$ is non-empty, at least for external
potentials satisfying the bounds (\ref{b1}) and (\ref{b2}).

In the non-rotating case, i.e., $\bfom=\0$, $\Engp(\0,g)$ and
$\Endm(\0,g)$ are equal for all $g\geq 0$. This remains true
if $|\bfom|$ is not too large. In fact one can show that there
exists an $\Omega_g>0$, depending on $g$, such that
$\Engp(\bfom,g)=\Endm(\bfom,g)$ for $|\bfom|\leq \Omega_g$ \cite{S02}.

Note that both $\Engp$ and $\Endm$ are concave functions of their
parameters and, in particular, continuous. Hence $\Xi$ is an open set,
being the complement of the zero set of the continuous function
$\Engp(\bfom,g)-\Endm(\bfom,g)$, and both axis $g=0$ and $|\bfom|=0$
are not contained in $\Xi$.

\section{Main Results}

With the preliminaries of the previous section in hand, we can now
state our main results. We are interested in the ground state
energy $\Eqm(N,\bfom,a)$ of the $N$-particle Hamiltonian
(\ref{ham}), for large $N$ and small $a$. In fact it turns out
that $a\sim N^{-1}$ is the case of interest. As explained in the
Introduction, $a\sim N^{-1}$ implies that all terms in the
Hamiltonian (\ref{ham}) yield a contribution to the ground state
energy of the same order as $N\to\infty$. That is, for some fixed
$g>0$, we will set $a=g N^{-1}$, or, more generally, we will
assume that $Na\to g$ as $N\to\infty$. Moreover, we will also
derive results for the corresponding ground states or, more
generally, for {\it approximate ground states}. We call a sequence
$\Psi_N\in \Hh_N$ an approximate ground state if $\|\Psi_N\|_2=1$
and
\beq\label{defappr}
\lim_{N\to\infty,\, Na\to g} \langle \Psi_N| H_{N,\bfomm,a} \,
\Psi_N\rangle \, \Eqm(N,\bfom, a) ^{-1} = 1
\eeq
for fixed $\bfom$ and $g$. Given such an approximate ground state,
we define its $n$-particle reduced density matrix by the kernel
\begin{multline}\label{defgamn}
\Gamma_N^{(n)}(\x_1,\dots,\x_n,\y_1,\dots,\y_n)= \frac 1{N!}
\sum_{\pi\in S_N} \int_{\R^{3(N-n)}} \prod_{j=n+1}^N d\x_j \\
\times
\Psi_N(\pi(\x_1,\dots,\x_N))
\Psi_N^*(\pi(\y_1,\dots,\y_n,\x_{n+1},\dots,\x_N))
\ ,
\end{multline}
where $S_N$ denotes the permutation group and $\pi(\x_1,\dots,\x_N)$
is a permutation of the $N$ variables $\x_i$, $1\leq i\leq N$. The
$^*$ denotes complex conjugation.  The $\Gamma_N^{(n)}$ defined in
(\ref{defgamn}) are trace class operators on $\Hh_n$, and the
normalization is chosen such that $\Tr[\Gamma_N^{(n)}]=1$. Our main
result on the properties of the absolute ground state of $\ham$ is the
following.

\begin{thm}[Ground state asymptotics]\label{T4}
For given $\bfom$ and $g$ let $\gdm_{\bfomm,g}$ be the unique
minimizer of $\Edm_{\bfomm,g}$, with corresponding energy
$\Endm(\bfom,g)$. Let $\Gamma_N^{(n)}$ denote the $n$-particle
reduced density matrix of an approximate ground state of
$H_{N,\bfomm,a}$. Then
\beq\label{upbol}
\lim_{N\to\infty}\frac 1N \Eqm(N,\bfom, g N^{-1})= \Endm(\bfom,g)
\eeq
uniformly in $g$ on compact intervals in $(0,\infty)$, and, for each
$n\in\N$,
\beq\label{34}
\lim_{N\to\infty} \Gamma_N^{(n)}= \underbrace{ \gdm_{\bfomm,g}
\otimes \dots \otimes \gdm_{\bfomm,g} }_{n {\rm \,\,  times}}
\eeq
in the usual norm of trace class operators on $\bigotimes_{i=1}^n
L^2(\R^3,d\x_i)$.
\end{thm}

This theorem will be proved in the next section. Concerning the
bosonic ground state energy $\Eqm_\bos(N,\bfom,g)$, we cannot give
the precise asymptotics as in (\ref{upbol}), but we can give upper
and lower bounds. One might conjecture that (\ref{upbos}) below
holds as equality, but we cannot prove this. However, the
conjecture is supported by the following theorem.

\begin{thm}[Asymptotics for bosonic ground state energy]\label{T5}
For fixed $\bfom$ and $g$ we have that
\beq\label{upbos}
\limsup_{N\to\infty}\frac 1N \Eqm_\bos(N,\bfom,gN^{-1})\leq
\Engp(\bfom,g) \ .
\eeq
Moreover,
\beq\label{36}
\liminf_{N\to\infty}\frac 1N \Eqm_\bos(N,\bfom,gN^{-1})>
\Endm(\bfom,g)
\eeq
if and only if
\beq\label{37}
\Engp(\bfom,g)>\Endm(\bfom,g) \ ,
\eeq
i.e., $(|\bfom|,g)\in \Xi$, and the same is true with $\liminf$
replaced by $\limsup$ in Eq.~(\ref{36}).
\end{thm}

\noindent {\bf  Remark 1.} Inequality (\ref{36}) holds uniformly in
$g$ for compact intervals of $g$ in~$\Xi$. More precisely, for any
$\eps>0$ such that the closed interval
$[(|\bfom|,g-\eps),(|\bfom|,g+\eps)]$ is contained in $\Xi$,
\beq\label{remark}
\liminf_{N\to\infty}
 \inf_{g'\in(g-\eps,g+\eps)}\left\{\frac 1N
\Eqm_\bos(N,\bfom,g'N^{-1})- \Endm(\bfom,g')\right\}>0 \ .
\eeq
This property will be important in the proof of
Corollary~\ref{C1}.

\bigskip

\noindent {\bf Remark 2.} 
Theorems~\ref{T4}
and~\ref{T5} together have the following consequences on the Bose
gas. If $(|\bfom|,g)\not\in \Xi$, e.g.,
$\Engp(\bfom,g)=\Endm(\bfom,g)$, there is a unique minimizer of
the GP functional (\ref{gpfunct}). Moreover,
\beq\label{qmqm}
\lim_{N\to\infty}\frac 1N E^{\rm
QM}_\bos(N,\bfom,gN^{-1})=\Engp(\bfom,g) \ ,
\eeq
as can be seen from the
lower bound (\ref{upbol}) and the upper bound (\ref{upbos}).
Therefore the bosonic ground state 
is an approximate ground state for the unrestricted
problem, and hence (\ref{34}) holds. The density matrix
$\gdm_{\bfomm,g}$ in this case is the one-dimensional projection
onto the minimizer of $\E_{\bfomm,g}$, and (\ref{34}) proves
complete {\it Bose-Einstein condensation} of all $n$-particle
density matrices. For $\bfom=\0$ this was proved in \cite{LS02}.
The persistence of BEC for $\bfom\neq \0$
can also be interpreted as a superfluid behavior of the system
(see \cite{LSYsfl}).

Note that these results are true not only for the bosonic ground
state, but for any approximate bosonic ground state. Moreover, the
notion of approximate ground states is readily generalized to
$N$-particle density matrices. In particular, the assertions above are
true for all Gibbs states
\beq\label{ggg}
\Gamma^\beta_{N,\bfomm,a} \equiv \frac {P_\bos \exp(-\beta \ham)}
{\Tr[P_\bos \exp(-\beta \ham)]} \ ,
\eeq
with $\beta>0$, assuming that the trace is finite, which is guaranteed
for external potentials $V$ that increase at least logarithmically in
$|\x|$ at infinity. We will show in Sect.~\ref{sect43} that
\beq\label{xxx}
\lim_{N\to\infty} \frac 1N\left( \Tr[H_{N,\bfomm,gN^{-1}} 
\Gamma^\beta_{N,\bfomm,gN^{-1}}]- E^{\rm QM}_\bos(N,\bfom,gN^{-1})\right)=0 
\eeq
for all $\beta>0$, implying, for $(|\bfom|,g)\notin\Xi$, complete BEC
of the reduced density matrices of (\ref{ggg}), i.e., they converge to
the right hand side of (\ref{34}), where $\gdm_{\bfomm,g}$ is now the
projection onto the unique GP minimizer. Note, however, that fixing
$\beta$ is really a {\it zero-temperature limit}, since the relevant
temperature scale depends on the (mean) density, which goes to
infinity in our limit. To obtain a true effect of the temperature one
has to scale it appropriately with~$N$. E.g., for a harmonic trap
potential the relevant temperature scale would be $T\sim N^{1/3}$
\cite{dalfovo}.

Note that for (\ref{xxx}) to hold true it is essential to restrict
oneselves to the bosonic subspace in (\ref{ggg}). Without this
restriction (\ref{xxx}) (with $E^{\rm QM}_\bos$ replaced by $E^{\rm
QM}$) will not be true, as can be seen from the non-interacting case
$g=0$.

\bigskip

A simple corollary of Theorem~\ref{T5} is the non-uniqueness
of ground states of the Hamiltonian $\ham$. Even more is true,
namely the ground state degeneracy grows exponentially with $N$.

\begin{cor}[Ground state degeneracy]\label{C1}
For $(|\bfom|,g)\in \Xi$ and $N$ large enough, no ground state of
$H_{N,\bfomm,gN^{-1}}$ has bosonic symmetry. Moreover, if
$\NN(N,|\bfom|,a)$ denotes the multiplicity of the ground state of
$\ham$, and $(|\bfom|,g)\in \Xi$, then
\beq
\liminf_{N\to\infty} \frac 1N \ln \NN(N,|\bfom|,gN^{-1})>0 \ .
\eeq
\end{cor}

\noindent {\bf Remark 3.} 
Corollary~\ref{C1} states that the degeneracy of the ground state of
$H_{N,\bfomm,gN^{-1}}$ grows {\em at least} exponentially with $N$ if
$(|\bfom|,g)\in\Xi$. If the external potential $V$ does not grow too
slowly at infinity, it is not difficult to see that in fact
$\NN(N,|\bfom|,gN^{-1})$ grows also {\em at most}
exponentially. Namely, if we assume that $\Tr[\exp(-\beta
H_0)]<\infty$ for some $\beta>0$, which is guaranteed if $V$ has at
least logarithmic increase in $|\x|$, then
\beq\label{rem2}
\limsup_{N\to\infty} \frac 1N \ln \NN(N,|\bfom|,gN^{-1})
\leq \ln \Tr\big[\exp\big(\beta(\Endm(\bfom,g)-H_0)\big)\big]  \ .
\eeq
This is shown in Sect.~\ref{44}.

\bigskip
The proofs of the results stated in this section can be found in
Section~\ref{proofsec} below. The energy difference of the bosonic
system to the one without symmetry restrictions can be understood as being
due to the positive correlation energy that gets added when trying to
symmetrize a state where {\it not} essentially all the particles
occupy the same state, as is the case in the regime of symmetry
breaking, where the minimizer of $\Edm$ has at least rank~2. It may
therefore be favorable for all bosons to occupy the same state,
restoring the complete BEC but breaking the rotational symmetry by
choosing one of the minimizers of the GP functional.

All the results in this paper refer to three-dimensional systems, but
analogous results can be obtained also for two-dimensional systems, as
has been shown, in the non-rotating case, in \cite{LSY01}. The DM and
GP functionals in this case have been studied in \cite{S02}. Moreover,
we could also allow for internal degrees of freedom of the particles,
e.g. spin, which would not affect the results on the absolute ground
state, but the ones on the bosonic ground state. In this case the GP
functional has to be replaced by the DM functional restricted to
density matrices of a definite rank equal to the number of internal
states (compare with Section~6 in \cite{S02}).

\section{Proofs}\label{proofsec}

Before we give the proof of the results stated in the previous
section, we state in Section~\ref{prel} two auxiliary Lemmas that 
will be used later. The proof of Theorem~\ref{T4} is given in
Section~\ref{sect42}, and the proof of Theorem~\ref{T5}, together
with the assertions made in Remarks~1 and~2, in
Section~\ref{sect43}. Finally, Corollary~\ref{C1} and the subsequent
Remark~3 are proved in Section~\ref{44}.

\subsection{Preliminaries}\label{prel}

The following Lemma is needed in the proof of Theorem \ref{T4}
below. It is related to the generalized Poincar\'e inequalities
studied in \cite{lsypoin}. For a measurable set $\Oset\subset
\R^3$ we denote by $\Oset^c$ its complement, and by $|\Oset|$ its
Lebesgue measure.

\begin{lem}\label{poinlem}
Let $\bfA\in L^2_{\rm loc}(\R^3;\R^3)$ and $V\in
L^\infty_{\rm loc}(\R^3;\R)$, and assume that
$\lim_{|\xx|\to\infty} V(\x)=\infty$. Let
\beq
E=\infspec \Big[ (-i\bfnab -\bfA(\x) )^2 + V(\x) \Big] \ ,
\eeq
and let $P$ denote the projector in $\Hh=L^2(\R^3,d\x)$ onto the
corresponding ground states. Let
\beq
\Delta E= \infspec  \big[(-i\bfnab -\bfA(\x) )^2 + V(\x) \big] 
 \restriction_{(1-P)\Hh} \, - \, E 
\eeq
denote the gap in the spectrum above the ground state energy, which is
positive because of the discrete spectrum of the operator under
consideration.

For all $\eps>0$ there exists a $\delta>0$ such that for all
$\Oset\subset \R^3$ with $|\Oset^c|<\delta$ and for all $f\in\Hh$
\begin{multline}
\eps \int_{\R^3} \left| (i\bfnab+\bfA)f\right|^2+ \int_\Oset
\left| (i\bfnab+\bfA)f\right|^2 + \int_{\R^3} V |f|^2\\ \geq E\,
\|f\|_{L^2(\R^3)}^2 + \Delta E\, \|f-P f\|_{L^2(\R^3)}^2 \ .
\end{multline}
\end{lem}

\begin{proof}
It is no restriction to assume that $V\geq 0$. As in
\cite{lsypoin} we will use a compactness argument. Suppose that
the Lemma is wrong. Then there exists an $\eps_0>0$ and a sequence
of pairs $(f_n,\Oset_n)$, such that
$\lim_{n\to\infty}|\Oset_n^c|=0$, $\|f_n\|_{L^2(\R^3)}=1$, and
\begin{multline}\label{cont}
\lim_{n\to\infty}\left[ \eps_0 \int_{\R^3} \left|
(i\bfnab+\bfA)f_n\right|^2+ \int_{\Oset_n} \left|
(i\bfnab+\bfA)f_n\right|^2\right. \\ \left. + \int_{\R^3} V
|f_n|^2 - \Delta E\, \|f_n-P f_n\|_{L^2(\R^3)}^2\right] \leq E \ .
\end{multline}
Now both $(i\bfnab+\bfA)f_n$ and $f_n$ are bounded sequences in
$L^2(\R^3)$, so we can pass to a subsequence that converges weakly
in $L^2(\R^3)$ to $(i\bfnab+\bfA)f$ and $f$, respectively. We may 
also assume that $\sum_n |\Oset_n^c|$ is finite. Defining $\Sigma_N$ by
$\Sigma_N=\R^3\setminus\bigcup_{n\geq N}\Oset_n^c$ we have
$\Sigma_N\subset\Oset_n$ for $n\geq N$. Using weak lower
semicontinuity of the norms in question, we therefore get
\begin{multline}\label{tog}
\liminf_{n\to\infty} \left[ \eps_0 \int_{\R^3} \left|
(i\bfnab+\bfA)f_n\right|^2+ \int_{\Oset_n} \left|
(i\bfnab+\bfA)f_n\right|^2 + \int_{\R^3} V |f_n|^2\right]
\\ \geq \sup_N \left[ \eps_0
\int_{\R^3} \left| (i\bfnab+\bfA)f\right|^2+ \int_{\Sigma_N}
\left| (i\bfnab+\bfA)f\right|^2 + \int_{\R^3} V |f|^2\right]
\\ = \left[ (1+\eps_0) \int_{\R^3} \left|
(i\bfnab+\bfA)f\right|^2+ \int_{\R^3} V |f|^2\right] > E \int_{\R^3}
|f|^2 \ .
\end{multline}
Now since $V$ goes to infinity at infinity, $-\Delta+V$ has a compact
resolvent (cf., e.g.,
\cite[Thm.\ XIII.65]{rs4}), and hence also 
$(i\bfnab+\bfA)^2+V$ has a compact resolvent
\cite[Thm.~2.7]{AHS}. Since
\beq
\int_{\R^3} \left( \left|(i\bfnab+\bfA)f_n \right|^2 + V|f_n|^2 \right)
< C
\eeq
for some $C<\infty$ independent of $n$, we can conclude that $f_n$ is
contained in a compact subset of $L^2(\R^3)$, and thus $f_n\to f$
strongly in $L^2(\R^3)$. This implies that $\|f\|_2 =1$, and also
\beq
\lim_{n\to\infty} \|f_n-Pf_n\|_{L^2(\R^3)} = \|f-Pf\|_{L^2(\R^3)}\
.
\eeq
Together with (\ref{tog}) this contradicts (\ref{cont}).
\end{proof}

Using more sophisticated methods, as in \cite{lsypoin}, it is
possible to investigate the relation between $\eps$ and $\delta$.
This is needed to get precise error estimates, but we shall not do
this here.

\bigskip
In the proof of Theorem~\ref{T4} it will be necessary to study a
homogeneous gas of $n$ particles, described by the Hamiltonian
\beq\label{homham}
- \sum_{i=1}^n \Delta_i + \sum_{1\leq i<j\leq n} \frac 1{a^2}
v((\x_i-\x_j)/a) \ ,
\eeq
acting on $L^2(\Lambda^n)$. The particles are confined to
$\Lambda=[0,L]^3$, a box of side length $L$, and we use {\it
Neumann} boundary conditions. A lower bound to the ground state
energy of (\ref{homham}) was obtained in \cite{LY98}, and we shall
describe the result here. In fact, the lower bound in \cite{LY98}
was obtained via a lower bound to the expression
\begin{equation}\label{49}
\sum_{i=1}^n \left[\frac\eps 2 \int_{\Lambda^n} |\bfnab_i f|^2 +
(1-\eps) \int_{K_i} |\bfnab_i f|^2\right] +
\sum_{i<j}\int_{\Lambda^n} \frac 1{a^2}v((\x_i-\x_j)/a)|f|^2 \ ,
\end{equation}
where $0<\eps<1$ and $K_i\subset\Lambda^n$ is given by
\beq
K_i=\left\{ (\x_1,\dots,\x_n)\in \Lambda^n \, : \, \min_{k,\,
k\neq i} |\x_i-\x_k|\leq R \right\} 
\eeq
for some $R>0$. This is exactly the expression that we
have to bound from below in the proof of Theorem~\ref{T4}, see Eq.
(\ref{ej2}). The result is the following. It is valid for all
spherically symmetric $v\geq 0$ with finite range and scattering
length $1$.

\begin{lem}\label{homlem}
Let $E_\eps(n,L)$ denote the infimum of (\ref{49}) over all
functions $f\in L^2(\Lambda^n)$ with $\|f\|_2=1$, and let $Y=a^3 n
/L^3$. If $R\geq a Y^{-5/17}$, $\eps\geq Y^{1/17}$, and $n\geq
Y^{-1/17}$, then there exists a constant $C>0$ such that
\beq\label{homlemeq}
E_\eps(n,L)\geq 4\pi a \frac{n^2}{L^3} \left( 1-C
Y^{1/17}\right)(1-\eps) \ .
\eeq
\end{lem}

The proof can be found in \cite{LY98} (see also \cite{LSSY} for a
more elaborate discussion). Strictly speaking, it was derived for
the expression (\ref{49}) with $\eps/2$ replaced by $\eps$ in
front of the first term, but it is easy to see that this
additional factor does not affect the main result, only the
constant appearing in (\ref{homlemeq}). Note that the condition on
$n$ means that $n\geq (L/ a)^{1/6}$.

\subsection{Proof of Theorem~\ref{T4}}\label{sect42}

We start by considering the case $n=2$ in (\ref{34}). In order to
be able to obtain information about the two-particle density
matrix, we start by introducing a modified DM functional. Let
$U\in C_0^\infty(\R^3\times\R^3)$ be realvalued and symmetric, i.e.,
$U(\x,\y)=U(\y,\x)$, and let $\delta\in \R$. The modified DM
functional is defined as 
\beq\label{dmmod}
\Edmmod[\gamma] =\Tr [ H_0\gamma] + 4\pi  g \int \rho_\gamma(\x)^2
d\x + \delta\int U(\x,\y) \rho_\gamma(\x)\rho_\gamma(\y) d\x d\y \
.
\eeq
We suppress the dependence on the parameters for simplicity of
notation. We assume that $\delta$ is small enough such that $\Edmmod$
is still strictly convex in $\rho_\gamma$, which is in particular the
case for $\delta \|U\|_{L^2(\R^6)} < 4\pi g$. The ground state energy
of (\ref{dmmod}) will be denoted by $\Enmod$, which depends, for fixed
$U$, on $\bfom$, $g$ and $\delta$.  As for the DM functional with
$\delta=0$, one can use standard methods to show the existence of a
minimizing density matrix for $\Edmmod$. Uniqueness is not clear,
however, but this is of no concern to us. (For $\delta$ small enough,
$\Edmmod$ is strictly convex in $\rho_\gamma$, and therefore the
density of a minimizer is unique. This convexity property will be
important in the proof of Lemma~\ref{l2} below.) Let $\gmod$ denote a
minimizer of $\Edmmod$, with corresponding density $\rmod$. One can
show that $\rmod$ is a bounded, continuously differentiable function
that decreases exponentially at infinity. Moreover, as in
Prop.~\ref{T1}, $\gmod$ also minimizes the linear functional
\beq
\gamma \mapsto \Tr[\Hmod \gamma] \ ,
\eeq
with $\Hmod$ given by
\beq\label{hmod}
\Hmod= H_0 + 8\pi g \rmod(\x) + 2 \delta \int U(\x,\y)\rmod(\y)
d\y \ .
\eeq
That is, the range of $\gmod$ is contained in the span of the ground
states of $\Hmod$, whose ground state energy is given by
\beq\label{infhmod}
\infspec \Hmod = \Enmod(\bfom,g,\delta)+ 4\pi g \int \rmod(\x)^2
d\x + \delta \int U(\x,\y) \rmod(\x) \rmod(\y) d\x d\y \ .
\eeq

It is important to note that $\Enmod$ is differentiable in
$\delta$ at $\delta=0$. This follows from concavity in $\delta$
and the fact that the minimizer at $\delta=0$ is unique (cf.,
e.g., \cite{LS77}). The derivative is given by
\beq\label{412}
\left. \frac{\partial \Enmod(\bfom,g,\delta)}{\partial \delta}
\right|_{\delta=0} = \int U(\x,\y) \rdm_{\bfomm,g}(\x)
\rdm_{\bfomm,g}(\y) d\x d\y \ .
\eeq

We will now evaluate upper and lower bounds on the ground state
energy of the $N$-particle problem. We also add our auxiliary
potential $U$ to the Hamiltonian, and estimate in the following the
quantity
\beq
E^{\rm QM}(N,\bfom, a,\delta) = \infspec \left[ H_{N,\bfomm,a} +
\frac{ 2\delta} N \sum_{1\leq i<j\leq N} U(\x_i,\x_j)\right] \ .
\eeq

We start with the upper bound. It is derived by analogous
considerations as in \cite{LSY00}, but instead of using a trial
vector we use an $N$-particle trial density matrix, whose kernel
has the form
\beq\label{kernel}
\Gamma(\x_1,\dots,\x_N,\y_1,\dots,\y_N)=
F(\x_1,\dots,\x_N)F(\y_1,\dots,\y_N)
\prod_{i=1}^N \gmod(\x_i,\y_i) ,
\eeq
where $\gmod$ is a minimizer of (\ref{dmmod}), and $F$ is the Dyson
wave function defined in \cite{LSY00}. It is given by
\beq\label{dyson}
F(\x_1,\dots,\x_N)=\prod_{i=1}^N f(t_i(\x_1,\dots,\x_i)) \ ,
\eeq
where $t_i = \min\{|\x_i-\x_j|, 1\leq j\leq i-1\}$ is the distance
of $\x_{i}$ to its nearest neighbor among the points
$\x_1,\dots,\x_{i-1}$, and $f$ is a function of $t\geq 0$. It is
chosen to be
\begin{equation}
f(t)=\left\{\begin{array}{cl} f_{0}(t)/f_0(b) \quad
&\mbox{for}\quad t<b\\ 1 &\mbox{for}\quad t\geq b\ ,
\end{array}\right.
\end{equation}
where $f_0$ is the solution of the zero energy scattering equation for 
the interaction potential, i.e., $-\Delta f+\half a^{-2} v(\x/a) f=0$,
and $b$ is some cut-off parameter of order $b\sim N^{-1/3}$. The
function $F$ is a suitable generalization of the function Dyson used
in \cite{dyson} to obtain an upper bound on the ground state energy of
a homogeneous Bose gas of hard spheres. The calculation of the upper
bound follows along the same lines as in
\cite{LSY00}, with the result that
\beq\label{up}
E^{\rm QM}(N,\bfom,gN^{-1},\delta)\leq N
\Enmod(\bfom,g,\delta)(1+O(N^{-2/3})) \ ,
\eeq
uniformly in $g$ on compact intervals.

We now proceed with the lower bound. Fix some $R>0$ and
$0<\eps<1$. They will be chosen later to depend on $N$ in a definite
way. We introduce the short hand notation
\beq
\X_i = (\x_1,\dots,\x_{i-1},\x_{i+1},\dots,\x_N)
\eeq
and
\beq
d\X_i = \prod_{ j=1,\, j\neq i}^N d\x_j \ .
\eeq
Moreover, let $\bfA(\x)=\half \bfom\times \x$, and let $r_i$
denote the radial distance of $\x_i$ orthogonal to $\bfom$. Let
$\bfnab_j = \bfnab_{\xx_j}$. Given any $\Psi\in\Hh_N$ with
$\|\Psi\|_2=1$, we can write the expectation value of our modified
Hamiltonian as
\beq\label{417}
\left\langle \Psi \left | H_{N,\bfomm,a}+ \frac {2\delta} N
\sum_{i<j} U(\x_i,\x_j) \right| \Psi \right\rangle = \sum_{j=1}^N
\left(E_j^{(1)} + E_j^{(2)}\right) \ ,
\eeq
where
\begin{multline}\label{ej1}
E_j^{(1)}= \int_{\R^{3(N-1)}} d\X_j \left[ (1-\eps)
\int_{\Oset_j^c} d\x_j \left|\left(i\bfnab_j+\bfA(\x_j)\right)
\Psi\right|^2 \right.
\\+ \left. \frac\eps 2 \int_{\R^3} d\x_j \left|
\left(i\bfnab_j+\bfA(\x_j)\right) \Psi\right|^2 \right.
\\ + \int_{\R^3} d\x_j \left( V(\x_j) - \frac{|\bfom|^2}4
r_j^2 + 8\pi g \rmod(\x_j) + 2\delta \int
U(\x_j,\y)\rmod(\y)d\y\right)|\Psi|^2 \Bigg]
\end{multline}
and
\begin{multline}\label{ej2}
E_j^{(2)}= \int_{\R^{3(N-1)}} d\X_j \left[ (1-\eps) \int_{\Oset_j}
d\x_j \left|\left(i\bfnab_j+\bfA(\x_j)\right) \Psi\right|^2
\right. \\ + \left. \frac\eps 2 \int_{\R^3} d\x_j
\left|\left(i\bfnab_j+\bfA(\x_j)\right) \Psi\right|^2 \right.
\\ \left. - \int_{\R^3} d\x_j \left( 8\pi g \rmod(\x_j) 
+ 2\delta \int U(\x_j,\y)\rmod(\y)d\y\right)|\Psi|^2\right. 
\\ \left. + \half \sum_{ i=1,\, i\neq j}^N \int_{\R^3} 
d\x_j \left( \frac 1 {a^{2}} v\left((\x_i-\x_j)/a\right) 
+ \frac {2\delta} N U(\x_i,\x_j) \right) |\Psi|^2  \right] \ .
\end{multline}
We choose, for fixed $\X_j$,
\beq
\Oset_j=\left\{ \x_j \in \R^3 \, : \, \min_{k,\, k\neq j}
|\x_j-\x_k|\leq R\right\} \ .
\eeq
In the following, we will investigate the two terms (\ref{ej1}) and
(\ref{ej2}) separately.  The results are formulated in the following
Lemmas. It is always understood that, for fixed $g$, $Na-g=o(1)$ as
$N\to\infty$, and that $\delta$ is small enough, as explained in the
beginning of Section~\ref{sect42}.

\begin{lem}\label{l1}
Let $\Pmod$ denote the projector onto the ground states of
$\Hmod$, defined in (\ref{hmod}). If $R\ll O(N^{-1/3})$ one can
choose $O(N^{-2/17})\ll \eps\leq o(1)$ as $N\to\infty$ such that
\begin{multline}
\sum_{j=1}^N E_j^{(1)} \geq N \left( \Enmod(\bfom,g,\delta)+ 4\pi
g \int \rmod(\x)^2 d\x\right. \\ \left. + \delta \int U(\x,\y)
\rmod(\x) \rmod(\y) d\x d\y+ C\, \Tr[\Gamma^{(1)}_\Psi (1-\Pmod)]
\right) (1-o(1))
\end{multline}
for some constant $C>0$ (depending on $\bfom$, $g$ and $\delta$).
Here $\Gamma_\Psi^{(1)}$ denotes the one-particle reduced density
matrix of $\Psi$.
\end{lem}

\begin{lem}\label{l2}
If $R\gg O(N^{-7/17})$ and $o(1)\geq \eps\gg O(N^{-2/17})$ as
$N\to\infty$, then
\beq
\sum_{j=1}^N E_j^{(2)} \geq - N  \left( 4\pi g \int \rmod(\x)^2
d\x + \delta \int U(\x,\y) \rmod(\x) \rmod(\y) d\x d\y
\right)(1+o(1)) \ .
\eeq
\end{lem}

Before proving these two Lemmas, let us show that they lead to
Theorem~\ref{T4}. Inserting the lower bounds to $\sum_j E_j^{(1)}$ and
$\sum_j E_j^{(2)}$ in (\ref{417}) we obtain, for an appropriate choice of
$R$ and $\eps$,
\begin{multline}
\frac 1N \left\langle \Psi \left | H_{N,\bfomm,a}+ \frac {2\delta}
N \sum_{i<j} U(\x_i,\x_j) \right| \Psi \right\rangle \\ \geq
\left[ \Enmod(\bfom,g,\delta)+ C\, \Tr[\Gamma^{(1)}_N(1-\Pmod)]
\right] (1-o(1))
\end{multline}
as $N\to\infty$, if $Na\to g$. Together with the upper bound
(\ref{up}) this implies that
\beq\label{star}
\lim_{N\to\infty} \frac 1 N E^{\rm QM}(N,\bfom,gN^{-1},\delta) =
\Enmod(\bfom,g,\delta)
\eeq
for all values of $\bfom$, $g$ and $\delta$ small, uniformly in
$g$ on compact intervals, and also that
\beq\label{stst}
\lim_{N\to\infty} \Tr[\Gamma_N^{(1)} (1-\Pmod)]=0 \ .
\eeq
Eq. (\ref{star}) for $\delta=0$ proves (\ref{upbol}). Moreover,
since $\Enmod$ is differentiable in $\delta$ at $\delta=0$, with
derivative given in (\ref{412}), we also infer that (see
\cite{LS77} for details)
\beq\label{details}
\lim_{N\to\infty} \int_{\R^6} U(\x,\y)\rho_N^{(2)}(\x,\y) d\x d\y
= \int_{\R^6} U(\x,\y) \rdm_{\bfomm,g}(\x) \rdm_{\bfomm,g}(\y) d\x
d\y \ ,
\eeq
where $\rho_N^{(2)}$ is the density of the reduced two-particle
density matrix $\Gamma_N^{(2)}$ of an approximate ground state at
$\delta=0$. This is true for all symmetric $U\in
C_0^\infty(\R^3\times \R^3)$, i.e., $\rho_N^{(2)}(\x,\y)$, being
itself symmetric, converges to
$\rdm_{\bfomm,g}(\x)\rdm_{\bfomm,g}(\y)$ in the sense of
distributions.

To show the convergence (\ref{34}) of the density matrices, not
only their densities, we proceed as follows. From (\ref{stst}) at
$\delta=0$ we infer that
\beq\label{pp}
\lim_{N\to\infty} \Tr[ \Gamma_N^{(2)} \Pdm\otimes\Pdm ] = 1 \ ,
\eeq
where $\Pdm$ is the projector onto the ground states of $\Hmod$
with $\delta=0$, which is the operator appearing in the linear
functional (\ref{deflin}). Note that $\Pdm$ is finite dimensional,
and $\Pdm \gdm_{\bfomm,g} \Pdm =  \gdm_{\bfomm,g}$ by
Prop.~\ref{T1}. From (\ref{pp}) we infer that there exists a
subsequence of $\Gamma_N^{(2)}$ that converges to some $\Gamma$ in
trace class norm, with $\Pdm\otimes\Pdm \Gamma
\Pdm\otimes\Pdm=\Gamma$.

We want to show that the convergence (\ref{details}) necessarily
implies that $\Gamma=\gdm_{\bfomm,g}\otimes \gdm_{\bfomm,g}$. To
do this, we have to take a closer look at the ground states of
$\Hmod$ for $\delta=0$. It is clear that they can be taken to be
eigenfunctions of $\bfom\cdot \bfL$, and that there is at most one
ground state for any given eigenvalue of $\bfom\cdot \bfL$. For
eigenvalue $m\in \Z$, the ground state wave function can be
written, in cylindrical coordinates $\x=(r,\varphi,z)$, as
\beq
f(r,z) e^{i m \varphi} \ ,
\eeq
with $f(r,z)>0$ for all $r>0$ and all $z$, and $f(r,z) r^{-|m|}$ bounded
(see \cite{S02} for details). Note that certainly $m\geq 0$ for any
ground state. Expanding $\Gamma$ in terms of these eigenfunctions, and
using the fact that because of (\ref{details}) its density is
necessarily given by $\rdm_{\bfomm,g}(\x)\rdm_{\bfomm,g}(\x')$, we
obtain the equation
\beq
\rdm_{\bfomm,g}(\x)\rdm_{\bfomm,g}(\x')= \sum_{j,k,l,m} a_{jklm}
f_j(r,z) f_k(r',z') f_l(r,z) f_m(r',z') e^{i(j-l)\varphi}
e^{i(k-m)\varphi'}
\eeq
for some coefficients $a_{jklm}$ that determine $\Gamma$. This sum is
finite, and since the left side does not depend on $\varphi$ and
$\varphi'$, we can infer that only terms with $j=l$ and $k=m$
contribute, i.e., $a_{jklm}=0$ for $j\neq l$ or $k\neq m$. Here we
have also used that terms with the same $j-l$, but different $j$ and
$l$ can not cancel, because of the different asymptotics as $r\to 0$,
namely $r^{j+l}$. By the same reasoning, we have
\beq
\rdm_{\bfomm,g}(\x)=\sum_j \lambda_j |f_j(r,z)|^2
\eeq
for some $0\leq\lambda_j\leq 1$. Using again the knowledge of the
$r\to 0$ asymptotics of the $f_j$, we infer that
$a_{jkjk}=\lambda_j \lambda_k$, and therefore
$\Gamma=\gdm_{\bfomm,g}\otimes\gdm_{\bfomm,g}$. This proves
(\ref{34}) for $n=2$.

It remains to prove Lemmas \ref{l1} and \ref{l2}.

\begin{proof}[Proof of Lemma \ref{l1}]
For fixed $\X_j$ define $f_j$ by
\beq
f_j(\x_j)=\Psi(\x_1,\dots,\x_j,\dots,\x_N) \ ,
\eeq
and let
\beq
W(\x)= V(\x)-\frac {|\bfom|^2}{4} r^2 + 8\pi g \rmod(\x)+ 2\delta
\int U(\x,\y)\rmod(\y) d\y \ .
\eeq
We have
\beq\label{ref}
E_j^{(1)}\geq (1-\eps) \int_{\R^{3(N-1)}} d\X_j F_j -
\frac\eps{1-\eps} \left\| \left[ W \right]_- \right\|_\infty \ ,
\eeq
with
\beq
F_j = \int_{\Oset_j} |(i\bfnab+\bfA)f_j|^2 + \frac \eps 2
\int_{\R^3} |(i\bfnab+\bfA)f_j|^2 + \int_{\R^3} W |f_j|^2 \ .
\eeq
The result now follows by applying Lemma \ref{poinlem} to $F_j$,
noting that $|\Oset_j^c|\leq N \mbox{$\frac {4\pi}3$} R^3 =
o(1)$ and that
\beq
\sum_{j=1}^N \int_{\R^{3(N-1)}} d\X_j \|f_j - P f_j\|_2^2 = N
\Tr[\Gamma_\Psi^{(1)}(1-\Pmod)]
\eeq
(compare with (\ref{hmod}) and (\ref{infhmod})). The constant $C$ is
then given by the spectral gap of $\Hmod$ above its ground state
energy.  Note that the last term in (\ref{ref}) is finite, by our
assumption (\ref{12}) on $V$ and the fact that $U$ is bounded and
$\rmod\in L^1(\R^3)$.
\end{proof}

\begin{proof}[Proof of Lemma \ref{l2}]
To obtain a lower bound, we can use the box method, as in
\cite{LY98,LSY00}. More precisely, we divide $\R^3$ into boxes of
side length $L$, labeled by $\alpha$, and distribute our $N$
particles over these boxes. Taking Neumann boundary conditions in
each box and minimizing the energy with respect to all
distributions of the particles, this can only lower the energy.

Let
\beq
W_\alpha = \sup_{\xx\in\alpha}\left[ 4\pi g \rmod(\x)+\delta
\int_{\R^3} U(\x,\y)\rmod(\y) d\y\right]
\eeq
and
\beq
U_{\alpha \beta} = \inf_{\xx\in\alpha,\, \yy\in\beta} U(\x,\y) \ .
\eeq
Using the diamagnetic inequality in the two terms in (\ref{ej2})
containing $\bfA$, we can set $\bfA=\bf0$ for a lower bound. With
$E_\eps(n,L)$ defined as in Lemma \ref{homlem}, we thus obtain
\begin{multline}\label{summ}
\sum_{j=1}^N E_j^{(2)} \geq\\ \inf_{\{n_\alpha\} }\left\{ \sum_\alpha
\left(E_\eps(n_\alpha,L) - 2 W_\alpha n_\alpha\right) + \frac
\delta {N} \sum_{\alpha,\beta} U_{\alpha \beta} n_\alpha n_\beta -
\frac \delta {N} \sum_\alpha U_{\alpha\alpha} n_\alpha\right\} \ ,
\end{multline}
where the infimum is taken over all distribution of the $N$
particles into the boxes $\alpha$. Each box contains $n_\alpha$ particles, and $\sum_\alpha n_\alpha=N$. The last term in (\ref{summ})
is due to the fact that there are only $\half n(n-1)$ pairs of
particles in the same box. It can easily be estimated by $\half
\delta \|U\|_\infty$, which is negligible compared to the other
terms of order $N$.

Let $\bar n_\alpha$ be a minimizing configuration of the $n_\alpha$'s
in (\ref{summ}). Let $Y=a^3 N/L^3$, and let $\Lambda_\eta$ denote the
collection of those boxes $\alpha$ where $\bar n_\alpha\geq \eta N
L^3$. If we choose $\eps\geq Y^{1/17}$ and $R\geq a (a^3 \eta
N)^{-5/17}$ (compare with the conditions stated in Lemma~\ref{homlem})
we can use (\ref{homlemeq}) to estimate, for $\alpha\in\Lambda_\eta$,
\beq
E_\eps(\bar n_\alpha,L)\geq 4\pi a\frac {\bar n_\alpha^2}{L^3}(1-C
Y^{1/17})(1-\eps) \ ,
\eeq
where we estimated $\bar n_\alpha$ by $N$ in the error term. For
$\alpha\not\in \Lambda_\eta$ we simply use $E_\eps(\bar
n_\alpha,L)\geq 0$. This gives
\begin{multline}\label{two}
(\ref{summ})\\ \geq \sum_{\alpha\in\Lambda_\eta,\, \beta\in\Lambda_\eta}
\bar n_\alpha \bar n_\beta \left[ \delta_{\alpha\beta} \frac{4\pi
g}{N L^3} (1-CY^{1/17})(1-\widehat\eps) + \frac{\delta} N
U_{\alpha\beta}\right] - \sum_{\alpha\in\Lambda_\eta} 2 W_\alpha \bar
n_\alpha \\- 2 \eta N L^3 \left(  \sum_\alpha
W_\alpha + \delta \sup_{\beta}\sum_\alpha U_{\alpha\beta}
\right)-\half \delta \|U\|_\infty \ .
\end{multline}
Here  $\widehat\eps = 1-(1-\eps)Na/g=o(1)$ for large $N$. 

In the following, we will choose $L=o(1)$ and $\eta=o(1)$ as
$N\to\infty$. The term in brackets in the last line in (\ref{two}) can
then be bounded by $\const L^{-3}$, which is, when multiplied by $\eta
N L^3$, of lower order than $N$ and therefore negligible compared to
the main terms of order $N$.

Consider now the first two terms in (\ref{two}). If $\delta$ is
small enough (independent of $N$ and $L$), the term in square
brackets defines a positive matrix $B_{\alpha\beta}$ for $N$ large
enough. Moreover, denoting
$\rho_\alpha=\sup_{\xx\in\alpha}\rmod(\x)$,
\beq
W_\alpha = NL^3 \sum_\beta B_{\alpha\beta} \rho_\beta +
\ell_\alpha \ ,
\eeq
where $\ell_\alpha=\ell_\alpha^{(1)}+\ell_\alpha^{(2)}$, with
\beq\label{ell1}
\ell_\alpha^{(1)}= 4\pi g
\left(CY^{1/17}+\widehat\eps-\widehat\eps C Y^{1/17}\right)
\rho_\alpha
\eeq
and
\beq\label{ell2}
\ell_\alpha^{(2)}= \delta \left( \sup_{\xx\in\alpha} \int
U(\x,\y)\rmod(\y)d\y-L^3 \sum_\beta
U_{\alpha\beta}\rho_\beta\right) \ .
\eeq
In the following we denote by $\matB$ the matrix with coefficients
$B_{\alpha\beta}$, and likewise by $\matW$, $\matrho$, $\matell$
the vectors with components $W_\alpha$, $\rho_\alpha$ and
$\ell_\alpha$, respectively. Let also $\matn$ denote the vector
with entries $\bar n_\alpha$ for $\alpha\in\Lambda_\eta$ and $0$ for
$\alpha\not\in\Lambda_\eta$. With these notations we can rewrite the
first two terms on the right side of (\ref{two}) as
\begin{multline}\label{450}
\langle \matn| \matB | \matn\rangle - 2 \langle \matW | \matn
\rangle
\\= \langle \matn-NL^3 \matrho | \matB |  
\matn-NL^3 \matrho\rangle -  \langle NL^3\matrho | 
\matB | NL^3 \matrho\rangle  - 2\langle \matell | \matn\rangle \ .
\end{multline}
Because of positivity of $\matB$, this expression is bounded below by
\begin{equation}
- \langle NL^3\matrho | \matB | NL^3 \matrho\rangle - 2 N
\|\matell\|_\infty \ .
\end{equation}
Now, if we choose $L=o(1)$ as $N\to\infty$, we have
\beq
(NL^3)^2 \langle \matrho | \matB | \matrho\rangle = N\left(4\pi
g\int \rmod(\x)^2 + \delta \int U(\x,\y)\rmod(\x)\rmod(\y) \right)
(1+o(1))
\eeq
because the sums are Riemann sums for the corresponding integrals.
Moreover, $\|\matell\|_\infty\leq o(1)$, which is obvious for
(\ref{ell1}), since $\rho_\alpha$ is bounded, and also for
(\ref{ell2}) by the same Riemann sum argument as above. This proves
the Lemma.
\end{proof}

This finishes the proof of (\ref{34}) for $n=2$. The general case
$n\geq 1$ follows in the same manner, perturbing the Hamiltonian
(\ref{ham}) by
\beq
\delta \frac {n!}{N^{n-1}} \sum_{1\leq i_1<\dots<i_n\leq N}
U(\x_{i_1},\dots,\x_{i_n}) \ ,
\eeq
where $U\in C_0^\infty(\R^{3n})$ is a realvalued symmetric function of $n$
variables, and also perturbing the DM functional by a term
\beq\label{term}
\delta \int_{\R^{3n}} U(\x_1,\dots,\x_n) \rho_\gamma(\x_1)\cdots
\rho_\gamma(\x_n) \ .
\eeq
We will only sketch the proof here, pointing out the differences
to the case $n=2$. Following the same steps as in the proof of
Lemma \ref{l2} above, one sees that minimizing over particle
numbers in different boxes is, after passing to the limit
$N\to\infty$, effectively equivalent to minimizing a functional
\beq\label{maps}
\rho \mapsto  \int_{\R^3} \left[4\pi g \rho(\x)^2 - W(\x)
\rho(\x)\right] + \delta \int_{\R^{3n}} U(\x_1,\dots,\x_n)
\rho(\x_1)\cdots \rho(\x_n)
\eeq
over functions $\rho\in L^1(\R^3)$ with $\rho\geq 0$ and $\int
\rho=1$. Here
\beq
W(\x)= 8\pi g \rmod(\x) + n \delta \int_{\R^{3(n-1)}}
U(\x,\x_2,\dots,\x_n) \rmod(\x_2)\cdots \rmod(\x_n) \ ,
\eeq
and $\rmod$ is the density of a minimizer of the modified DM
functional with additional term (\ref{term}), with $\int\rmod=1$.
To obtain the analogue of Lemma \ref{l2} we have to to show that
the minimizer of (\ref{maps}) is given by $\rmod$, at least for
$\delta$ small enough (compare with (\ref{450}), which is just a
discretized version of (\ref{maps}) for $n=2$). Compared to the
case $n=2$ there is an additional difficulty here since the
functional (\ref{maps}) is not necessarily convex. (The convexity
in the case $n=2$ was used in the form of positivity of $\matB$.)

To show that, for small enough $\delta$, the minimizer of the
functional (\ref{maps}) is given by $\rmod$, we rewrite
(\ref{maps}) as
\beq
-4\pi g \int \rmod(\x)^2 - (n-1)\delta \int
U(\x_1,\dots,\x_n)\rmod(\x_1)\cdots \rmod(\x_n) + \F[\rho] \ ,
\eeq
with
\begin{multline}
\F[\rho] = 4\pi g\int(\rho-\rmod)^2 \\ + \sum_{i=1}^n c_i
\int_{\R^{3i}}
U^{(i)}(\x_1,\dots,\x_i)(\rho(\x_1)-\rmod(\x_1))\cdots
(\rho(\x_i)-\rmod(\x_i)) \ ,
\end{multline}
where
\beq
U^{(i)}(\x_i,\dots,\x_i)=\int_{\R^{3(n-i)}}
U(\x_1,\dots,\x_n)\rmod(\x_{i+1})\cdots\rmod(\x_n)d\x_{i+1}\cdots
d\x_n
\eeq
and $c_i\in\N$ are appropriate integer coefficients. Now
$\F[\rmod]=0$, so any minimizer $\rho$ must fulfill $\F[\rho]\leq 0$.
Since $\|\rho-\rmod\|_1\leq 2$ by assumption, we have
$\F[\rho]\geq 4\pi g \|\rho-\rmod\|_2^2 - \const \delta
\|U\|_\infty$, and hence
\beq\label{henc}
\|\rho-\rmod\|_2^2 \leq \const \delta
\eeq
for a minimizer $\rho$. On the other hand,
\beq
\F[\rho]\geq 4\pi g\|\rho-\rmod\|_2^2 - \const \delta \|U\|_2
\left( \|\rho-\rmod\|_2^2 + \|\rho-\rmod\|_2^n \right) \ ,
\eeq
which is, because of (\ref{henc}), positive for $\delta$ small
enough, and zero only for $\rho=\rmod$. Therefore $\rmod$ is the
unique minimizer of (\ref{maps}).

The convergence of the energies for a small interval of $\delta$'s
around zero implies again the convergence of the $n$-particle
densities at $\delta=0$, and by analogous arguments as in the case
$n=2$ one can show the convergence of the density matrices.  The
details are left to the reader.

This finishes the proof of Theorem~\ref{T4}. 

\subsection{Proof of Theorem~\ref{T5}}\label{sect43}

We start by computing an upper bound to $E^{\rm QM}_\bos$. Note that
we can not use (\ref{kernel}) as a trial state, for two reasons. First
of all, $F$ is not a symmetric function and, secondly, the tensor
product of one-particle density matrices is not a bosonic density
matrix, if their rank is bigger than one. The first problem turns out
not be serious (see below), but the second is.

Instead of (\ref{kernel}), we use as a trial function
\beq\label{psibos}
\Psi(\x_1,\dots,\x_N)=F(\x_1,\dots,\x_N)\prod_{i=1}^N \phi^{\rm
GP}(\x_i)\ ,
\eeq
where $\phi^{\rm GP}$ is a minimizer of the GP functional, and $F$
is given by the Dyson wave function (\ref{dyson}). Note that $F$
is not a symmetric function, but we claim that (\ref{psibos})
nevertheless gives an upper bound to $\Eqm_\bos$. More precisely,
we will show that any trial function of the form (\ref{psibos})
with $F$ {\it realvalued} but not necessarily symmetric gives an
upper bound to $\Eqm_\bos$. To see this, we write, for $\Psi$
given in (\ref{psibos}),
\begin{multline}\label{mmm}
\left\langle \Psi \left| \ham \right|\Psi\right\rangle = \mu^{\rm
GP}_{\bfomm,g}\langle\Psi|\Psi\rangle+ \left. \sum_{j=1}^N
\int_{\R^{3N}} \prod_{k=1}^N |\phi^{\rm GP}(\x_k)|^2 d\x_k \right[
|\bfnab_j F|^2  \\ + \left.  \left( \half \sum_{i=1,\, i\neq j}^N
\frac 1{a^2} v((\x_i-\x_j)/a) - 8\pi g |\pgp(\x_j)|^2 \right)
|F|^2 \right] \prod_{k=1}^N |\phi^{\rm GP}(\x_k)|^2 d\x_k \ ,
\end{multline}
where we used partial integration, the GP equation (\ref{gpeq}),
and the fact that $F$ is real by assumption. The reality of $F$
has the effect the both the cross-term in the partial integration
and the term with $\bfL$ acting on $F$ vanish. The infimum of
(\ref{mmm}) over all $F$ is attained for a positive $F$, and
therefore we can proceed as in \cite[p.~15]{L90} to show that the
infimum of (\ref{mmm}) over all $F$ is the same as the infimum
over symmetric $F$.

It is therefore legitimate to use (\ref{psibos}) as a trial function
for the bosonic ground state problem. The calculation of the
expectation value of $\ham$ follows again along the same lines as
in \cite{LSY00}, with the result that
\beq
E^{\rm QM}_\bos(N,\bfom,gN^{-1})\leq N
\Engp(\bfom,g)(1+O(N^{-2/3})) \ .
\eeq
Letting $N\to\infty$ this proves (\ref{upbos}).

From (\ref{upbos}) it follows immediately that
\beq\label{366}
\limsup_{N\to\infty}\frac 1N \Eqm_\bos(N,\bfom,gN^{-1})>
\Endm(\bfom,g)
\eeq
implies (\ref{37}). Since (\ref{36}) clearly implies (\ref{366}), it
remains to show that (\ref{36}) follows from (\ref{37}). To see this,
note that $\Engp(\bfom,g)> \Endm(\bfom,g)$ implies that the unique
minimizer $\gdm_{\bfomm,g}$ of $\Edm_{\bfomm,g}$ has at least rank
$2$. Therefore $\gdm_{\bfomm,g}\otimes \gdm_{\bfomm,g}$ is {\it not} a
bosonic two-particle density matrix, it has both symmetric and
antisymmetric eigenfunctions. Since the space of bosonic two-particle
density matrices is closed under the trace norm, (\ref{34}) cannot be
true for the bosonic ground state in this case. Theorem~\ref{T4}
implies that the bosonic ground state is not an approximate ground
state for the unrestricted problem, and hence (\ref{36}) holds. This
proves Theorem~\ref{T5}.  The uniformity of the limit (\ref{34}) (see
the definition of approximate ground state in (\ref{defappr})) implies
uniformity in (\ref{36}), as stated in (\ref{remark}) in Remark~1.

We conclude this section with a proof of the assertion (\ref{xxx}) 
made in Remark~2 that the 
Gibbs states (\ref{ggg}) are approximate ground states. 
Let $T=1/\beta$. We start by
estimating the {\it free energy}
\beq
F(T,N)=-\frac 1\beta \ln \Tr[P_\bos \exp(-\beta H_N)] \ .
\eeq
For simplicity of notation we suppress the dependence on $\bfom$ and
$a$ from now on. It is always understood that $a=gN^{-1}$. The
function $F$ is monotone decreasing and concave in $T$. This implies
immediately that $F(T,N)\leq F(0,N)= E^{\rm QM}_\bos(N)$. To obtain a
lower bound, we proceed as follows. Let $n(e)$ denote the number of
eigenvalues of $H_N\restriction_{P_\bos \Hh_N}$ that are smaller or
equal to $e$. Then
\beq
F(T,N)=-\frac 1\beta \ln\left[\beta \int_{E^{\rm QM}_\bos(N)}^\infty
n(e)\exp(-\beta e) de \right] \ .
\eeq
Since $v\geq 0$, $n(e)$ is bounded above by the number of eigenvalues
of the non-inter\-acting Hamiltonian $\sum_i
H_0^{(i)}\restriction_{P_\bos
\Hh_N}$ that are smaller or equal to $e$. This, in turn, can be
bounded above by
\beq
\Tr\big[P_\bos \exp\big(\alpha e-\alpha \mbox{$\sum_i$} 
H_0^{(i)}\big)\big] \equiv \exp(\alpha e) C_{\alpha,N}
\eeq
for any $\alpha>0$. Choosing $\alpha<\beta$ we therefore obtain
\beq\label{fff}
F(T,N)\geq \frac{\beta-\alpha}\beta E^{\rm QM}_\bos(N) - \frac 1\beta
\ln\left(\frac{\beta C_{\alpha,N}}{\beta-\alpha}\right) \ .
\eeq
Now let $e_i$, $i\geq 0$, denote the ordered sequence of eigenvalues
of $H_0$, including degenerate eigenvalues, and let $j_0$ denote the
multiplicity of the lowest eigenvalue $e_0$. The expression
$C_{\alpha,N}$ can easily be bounded above using the fact that any
totally symmetric eigenfunction of $\sum_i H_0^{(i)}$ can be
characterized by the occupation numbers $n_j$ of the eigenfunctions of
$H_0$ corresponding to the eigenvalue $e_j$. Thus $C_{\alpha,N}$ can be expressed as 
\begin{eqnarray}\nonumber
C_{\alpha,N}&=&e^{-\alpha N e_0} \Tr \left[ P_\bos \exp\left(-\alpha
\mbox{$\sum_{i=1}^N$}\big( H_0^{(i)}-e_0\big)\right) \right] \\
 &=& e^{-\alpha N e_0} \sum_{\{n_j\in\N_0\}_{ j\geq 0} , \, \sum
n_j=N} e^{-\alpha \sum n_j(e_j-e_0)} \ ,
\end{eqnarray}
where we introduced the notation $\N_0=\N\cup \{0\}$. To bound this
expression from above, we simply neglect the condition $\sum_{j\geq 0}
n_j = N$. More precisely, we estimate
\begin{eqnarray}\nonumber
C_{\alpha,N}&\leq&e^{-\alpha N e_0}
\sum_{n_0=0}^N \cdots \sum_{n_{j_0-1}=0}^N\quad  
\sum_{\{n_j\in\N_0\}_{j\geq j_0} } e^{-\alpha \sum n_j(e_j-e_0)}\\ \label{fini}
&=& (N+1)^{j_0} 
e^{-\alpha N e_0} \prod_{j=j_0}^\infty \frac 1{1-e^{-\alpha(e_j-e_0)}}
\ .
\end{eqnarray}
Note that the last product is finite, and independent of $N$. 
Inserting the bound (\ref{fini}) in (\ref{fff}), letting $N\to\infty$ and
afterwords $\alpha\to 0$, we obtain
\beq\label{asdf}
\lim_{N\to\infty} \frac 1N \left( F(T,N) - E^{\rm QM}_\bos(N)\right)=0 \ . 
\eeq
To see that the same convergence holds for
the energies
\beq 
E(T,N)= \frac {\Tr[P_\bos H_N \exp(-\beta H_N)] }{\Tr[P_\bos
\exp(-\beta H_N)] } \ ,
\eeq
note that 
\beq
\frac {\partial F(T,N)}{\partial T} = \frac 1 T\big(F(T,N)-E(T,N)\big) \ .
\eeq
The concavity of $F(T,N)$ in $T$ and (\ref{asdf}) imply that
$N^{-1} \partial F/\partial T \to 0$ as $N\to\infty$, which proves the
desired result.

\subsection{Proof of Corollary \ref{C1}}\label{44}

The first statement on the impossibility of having a totally symmetric
ground state follows immediately from Theorem~\ref{T5}. To estimate
the ground state degeneracy, we proceed as follows. Since the
Hamiltonian under consideration is symmetric with respect to exchange
of two variables, any ground state belongs to a finite dimensional
space of an irreducible representation of $S_N$, the permutation group
of $N$ variables. These representations are conveniently labeled by
Young tableaux. Given the ground state space of $H_{N,\bfomm,gN^{-1}}$
and one of the corresponding representations of $S_N$, let, for fixed
$(|\bfom|,g)\in\Xi$, $R_N$ and $C_N$ denote the length of the longest
row and column of the Young tableaux, respectively. We will show that
$C_N$ is bounded above independent of $N$, and $R_N/N<1-\eps$ for some
$\eps>0$ and $N$ large enough. This will allow us to show that the
dimension of the representation increases exponentially with $N$.

We first compute an upper bound on $C_N$. There exists a ground
state of $H_{N,\bfomm,gN^{-1}}$ which is antisymmetric in the
first $C_N$ variables. Neglecting the interaction with the other
particles, we therefore have
\beq
E^{\rm QM}(N,\bfom,gN^{-1})\geq \Tr [ H_0 P_{C_N}] + E^{\rm
QM}(N-C_N,\bfom,gN^{-1}) \ ,
\eeq
where $P_C$ is the projection of $H_0$ onto the first $C$
eigenvalues. By analogous considerations as in the upper bound to
$E^{\rm QM}$ (cf. also \cite[Thm.~III.2]{LSY00}) it is not
difficult to show that
\beq
E^{\rm QM}(N,\bfom,gN^{-1})-E^{\rm QM}(N-C,\bfom,gN^{-1})\leq
\const C
\eeq
for fixed $g$ and $\bfom$, independent of $N$. Since
\beq
\lim_{C\to\infty} \frac 1C  \Tr [ H_0 P_C] = \infty
\eeq
because of our assumption that $V$ goes to infinity at infinity,
this implies that $C_N$ has to stay bounded as $N\to\infty$.

We now derive an upper bound on $R_N$. Since there is a ground
state of $H_{N,\bfomm,gN^{-1}}$  which is symmetric in the first
$R_N$ variables, we have, by analogous considerations as above,
\beq\label{468}
E^{\rm QM}(N,\bfom,gN^{-1})\geq E^{\rm QM}_\bos(R_N,\bfom,gN^{-1})
+(N-R_N) E_0 \ ,
\eeq
where $E_0=\infspec H_0$. Let $\eta=\limsup_{N\to\infty}R_N/N$,
and  choose a subsequence such that $R_N/N\to\eta$ as
$N\to\infty$. It follows from (\ref{468}) and Theorem~\ref{T4}
that
\beq
\eta \leq
\left(\Endm(\bfom,g)-E_0\right)\left(\liminf_{N\to\infty}\frac
1{R_N} E^{\rm QM}_\bos(R_N,\bfom,gR_N^{-1}\, R_N/N))-E_0\right)^{-1}
\ .
\eeq
Since certainly $R_N\to\infty$ with $N$, we can use (\ref{remark})
to conclude that
\beq
\liminf_{N\to\infty}\frac 1{R_N} E^{\rm
QM}_\bos(R_N,\bfom,gR_N^{-1}\, R_N/N)) > \Endm(\bfom,g\eta)
\eeq
if $1-\eta$ is small enough such that $(|\bfom|,g\eta)\in \Xi$.
This shows that $\eta<1$ for $(|\bfom|,g)\in\Xi$.

We have thus shown that, for some $\eps>0$, $R_N\leq N(1-\eps)$ and
$C_N\leq 1/\eps$ for $N$ large enough. Therefore there are at least
two rows of length $\geq \eps^2 N$ in the Young tableaux, and hence
the dimension of the corresponding representation is at least the one
of the tableaux consisting of two rows of length $\ell=[\eps^2 N]$
(here the square bracket denotes the integer part). This is given by
\beq
\frac {(2\ell)!}{\ell ! (\ell+1)! } \sim \frac
{4^\ell}{\ell^{3/2}} \quad {\rm for\ large\ }\ell,
\eeq
which proves the Corollary.

To prove the statement on the upper bound on the ground state
degeneracy in Remark~3 after Cor.~\ref{C1} we estimate, for
$\theta(z)=1$ for $z\geq 1$ and $\theta(z)=0$ for $z<1$,
\begin{eqnarray}\nonumber
\NN(N,|\bfom|,gN^{-1})&=&\Tr\big[ \theta\big(
E^{\rm QM}(N,\bfom,gN^{-1})-H_{N,\bfomm,gN^{-1}}\big)\big] 
\\ &\leq& \Tr\big[\theta\big( N(\Endm(\bfom,g)+\eps) 
- \mbox{$\sum_{i=1}^N$} H_0^{(i)}\big)\big]
\end{eqnarray}
for any $\eps>0$ and $N$ large enough. Here we have used (\ref{upbol})
and the fact that the interaction potential $v$ is positive by
assumption. Estimating the step function by the exponential function,
we therefore get, for any $\beta>0$, 
\begin{eqnarray}\nonumber
\NN(N,|\bfom|,gN^{-1})&\leq &
\Tr\big[ \exp\big(\beta( N(\Endm(\bfom,g)+\eps) - 
\mbox{$\sum_{i=1}^N$} H_0^{(i)})\big)\big] \\ 
&=&  \Tr\big[ \exp\big(\beta( \Endm(\bfom,g)+\eps - H_0)\big)\big]^N \ ,
\end{eqnarray}
which proves (\ref{rem2}).

\appendix
\section{Appendix: Proof of Theorem~\ref{T3}}

The strategy of the proof of Theorem~\ref{T3} is similar to the one in
the two-dimensional case given in in \cite{S02}.  One shows the
non-uniqueness of minimizers of the GP functional by showing that no
eigenfunction of the angular momentum \mbox{$\bfom\cdot\bfL$} can be a
minimizer, even though the GP functional is invariant under rotations
around the $\bfom$ axis. This implies that the minimizer cannot be
unique, since by rotating a minimizer one gets another one. Moreover,
since the minimizer of the DM functional $\Edm_{\bfomm,g}$ is unique
by Prop.~\ref{T1}, this implies that it has to have at least rank 2,
and $\Endm(\bfom,g)<\Engp(\bfom,g)$, as stated in Theorem~\ref{T3}.

Thus all we have to show is that an eigenfunction of $\bfom\cdot\bfL$
cannot minimize $\E_{\bfomm,g}$. To do this, we first have to consider
the properties of minimizers in the subspaces of angular momentum
eigenfunctions. Except for Lemma~\ref{la2}, all the considerations
below follow the same line as in the two-dimensional
case. Lemma~\ref{la2}, however, is the main new ingredient to extend
the results to three dimensions. While Lemmas~\ref{la1}--\ref{la3}
hold for general $V$'s, our assumptions (\ref{b1}) and (\ref{b2}) on
the external potential $V$ will become important in the proof of
Theorem~\ref{T3} below, where they are used in order to get explicit
estimates on the various quantities appearing in
Lemmas~\ref{la1}--\ref{la3}.

Let $f_n$ denote the minimizer of the GP functional restricted to
functions $\phi(\x)=f(r,z) e^{i n \varphi}$, i.e., to functions
with angular momentum $\bfom\cdot\bfL= n$. More precisely, $f_n$
minimizes
\beq\label{En}
f\mapsto  \int_0^\infty d^2 r \int_{-\infty}^\infty dz f^* \left(
-\Delta_r f -\frac{\partial^2 f}{\partial z^2}+\frac {n^2}{r^2}
f+V f + 4\pi g |f|^2 f \right) 
\eeq
under the normalization condition $ \int
|f(r,z)|^2 d^2 r dz =1$, where we denoted $d^2r=2\pi r dr$, and
$\Delta_r=\partial^2/\partial r^2+r^{-1} \partial/\partial r$. In the
following, it will be convenient to study this functional for all
$n\in\R_+$, not only for integers. Using standard methods one can
show that for each $n\geq 0$ and $g\geq 0$ there exists a unique
minimizer of (\ref{En}). Let $E_n(g)$ denote the corresponding
minimum of (\ref{En}), and let
\beq
\widetilde\mu_n = E_n(g) + 4\pi g \int |f_n(r,z)|^4  d^2r dz \ .
\eeq
Note that $\widetilde\mu_n$ depends on $g$ besides $n$.

We need the following estimates on the minimizers $f_n$.

\begin{lem}\label{la1}
The minimizer $f_n$ is a non-negative, bounded function, with
\beq\label{est1}
\|f_n\|_\infty \leq \frac{\widetilde\mu_n}{8\pi g} \ .
\eeq
\end{lem}

\begin{proof}
As in \cite[Lemma 1]{S02} this follows from the maximum principle,
applied to the variational equation for $f_n$, which reads
\beq
-\Delta_r f_n - \frac{\partial^2}{\partial z^2} f_n
+\frac{n^2}{r^2}f_n + Vf_n =\left( \widetilde\mu_n - 8\pi g
|f_n|^2\right) f_n \ .
\eeq
Recall that $V\geq 0$ by assumption. 
\end{proof}

\begin{lem}\label{la2}
Let $j_n(r)=\int_{-\infty}^\infty dz |f_n(r,z)|^2$. Choose $R>0$
large enough such that
\beq\label{defdel}
\delta_n(R)\equiv \inf_{|z|\geq R} V(\x) - \widetilde\mu_n>0 \ ,
\eeq
which is always possible by our assumption that $V$ goes to infinity
at infinity. Then
\beq\label{est2}
\|j_n\|_\infty \leq \|f_n\|_\infty^2 \left( 4R  + \const \frac
1{R\delta_n(R)} \right) \ .
\eeq
Here $\|\, \cdot \, \|$ has to be understood as the supremum norm
on $\R^2$ and $\R^3$, respectively.
\end{lem}

\begin{proof}
Let $\eta\in C^\infty(\R)$ be such that $0\leq \eta\leq 1$,
$\eta(z)=1$ for $|z|\geq 2$ and $\eta(z)=0$ for $|z|\leq 1$. Then
\beq
j_n(r)= \int_{-\infty}^\infty dz |f_n(r,z)|^2 \big( (1-\eta(z/R)) +
\eta(z/R) \big) \leq 4 R \|f_n\|_\infty^2 + \chi(r) \ ,
\eeq
where we denoted 
\beq
\chi(r)= \int dz |f_n(r,z)|^2 \eta(z/R) \ .
\eeq
Using the variational equation for $f_n$ and neglecting positive
terms, it is straightforward to derive the inequality
\beq\label{a9}
-\Delta_r \chi \leq 2 \int dz \eta(z/R) \left(|f_n|^2
\left(\widetilde\mu_n - V(r,z)  \right) + f_n
\frac{\partial^2}{\partial z^2} f_n \right) \ .
\eeq
Partial integration yields the estimate
\beq
2 \int dz \eta(z/R) \left( f_n \frac{\partial^2}{\partial z^2} f_n
\right)\leq \frac 1{R^2} \int dz |f_n|^2 \eta''(z/R)\leq \frac
\const R \|f\|_\infty^2\equiv \eps  \ .
\eeq
Using (\ref{defdel}) and the support properties of $\eta$ we
therefore get
\beq\label{a11}
-\Delta_r \chi \leq \eps - 2\delta_n(R) \chi\ .
\eeq
It follows from the maximum principle that $\chi(r)\leq
\eps/(2\delta_n(R))$, which proves our claim.
\end{proof}

\begin{lem}\label{la3}
With $j_n$ given in Lemma~\ref{la2}, $r^{-2n} j_n(r)$ is a bounded
function, with
\beq\label{lll}
\|r^{-2n} j_n\|_\infty \leq \left(2 c_n^2 \widetilde\mu_n\right)^n
\|j_n\|_\infty \ ,
\eeq
where $c_n$ is given by
\begin{eqnarray} \nonumber
&&c_n=\left(\frac{2^{-n}\left(\frac{2-n}n\right)^{n/2}\pi\,{\rm
Csc}\left(\frac{n\pi}2\right)} {(2-n)\Gamma(n)} \right)^{1/n}
\quad\mbox{for $n\leq 1$} \\ \label{cn}
&&c_n=\frac{\sqrt\pi}n\frac{\Gamma(n+\half)} {\Gamma(n)}
\quad\mbox{for $n\geq 1$\ .}
\end{eqnarray}
\end{lem}

\begin{proof}
The proof of this Lemma is similar to the corresponding Lemma~2 in
\cite{S02}. Let $h(r)=r^{-2n} \chi(r)$. Using the variational
equation for $f_n$ and the fact that $V$ is positive by
assumption, one derives the differential inequality
\beq\label{a14}
-h''(r)-\frac {(2n+1)}r h'(r) \leq 2\widetilde\mu_n h(r) \ .
\eeq
Multiplying (\ref{a14}) with the integral kernel of the resolvent of
the operator on the left side of this inequality and integrating, we
can proceed as in \cite[Lemma~2]{S02} to arrive at (\ref{lll}).
\end{proof}

We now are able to give the proof of Theorem~\ref{T3}.

\begin{proof}[Proof of Theorem~\ref{T3}]
We start by estimating the difference of $E_{n+1}(g)$ and
$E_n(g)$. By analogous considerations as in \cite[Eqs.~(3.2) and
(3.7)--(3.11)]{S02} we have, using Lemma~\ref{la3} in replacement
of \cite[Lemma~2]{S02},
\begin{multline}
E_{n+1}(g)-E_n(g)\leq (2n+1)\left(E_{1}(g)-E_0(g)\right)
\\ \leq (2n+1)2\pi e  \sup_{0<m<1} \|j_m\|_\infty
\max\left\{3, 1+\ln\left(\frac{\widetilde\mu_m}{2\pi\|j_m\|_\infty}\right)\right\} \ .
\end{multline}
We now use the bounds (\ref{est2}) and (\ref{est1}) in order to
estimate $\|j_m\|_\infty$. Our assumptions on the external
potential $V$ imply that, if $\delta_n(R)=O(1)$,
\beq
\mu_n \sim g^{1/(1+2/s+1/p)} \qquad {\rm and\qquad } R\sim \mu_n^{1/p}
\eeq
for large $g$, uniformly in $n$ for bounded $n$. Therefore
\beq\label{ennn}
E_{n+1}(g)-E_n(g)\leq O\left( g^{-\mbox{$\frac 2s$}(2+2/s+1/p)}\ln
(g)\right)\xrightarrow{g\rightarrow\infty} 0 \ .
\eeq
Note that the minimum of the GP functional over all functions with
angular momentum $n$ is given by $E_n(g)-n|\bfom|$, and hence, for
any fixed given $N$, all functions with angular momentum $< N$
cannot be GP minimizers if
\begin{equation}
\min_{0\leq n<N}\{E_n(g)-n|\bfom|\}>\min_{n\geq N}
\{E_n(g)-n|\bfom|\} \ .
\end{equation}
This is fulfilled if
\begin{equation}\label{aha}
|\bfom|>\max_{0\leq n<N}\frac{E_N(g)-E_n(g)}{N-n} \ ,
\end{equation}
which holds true, for any given fixed $N$ and $\bfom\neq \0$, for
large enough $g$ by (\ref{ennn}).

It remains to show that for given $\bfom$ there exists an
$N_{\bfomm}$, independent of $g$, such that all functions with
angular momentum $\geq N_{\bfomm}$ cannot be GP minimizers. We do
this by showing that they are unstable, in the sense that a small
perturbation lowers the energy. This can be done by a similar
calculation as in the two-dimensional case \cite{S02}, using the
estimates of Lemmas~\ref{la1}--\ref{la3} above. We will only
sketch the argument here, mainly pointing out the difference to
the two-dimensional case.

In order to show instability, one has to show that the second
derivative of the GP functional at some given stationary state
$\phi(\x)=f_n(r,z)e^{in\varphi}$ is negative in some direction.
More precisely,
\beq\label{aaa}
Q(w)\equiv\left\langle w\left| H_0 + 16\pi g |f_n|^2 -
\widetilde\mu_n+n|\bfom| \right|w\right\rangle + 8\pi g \,\Re \int
\phi^{*2} w^2 < 0
\eeq
for some $w$ with $\langle\phi|w\rangle=0$. Here $\Re$ denotes the
real part of a complex number. We will choose an $w$ that, when added
to the function $f_n(r,z)e^{in\varphi}$, effectively splits the
central vortex of degree $n$ into $n$ different vortices of degree $1$
that are arranged symmetrically around the axis $r=0$. The idea is
that for large enough $n$ a splitting of a central vortex like this
should lower the energy of the GP functional, and therefore the second
derivative given by $Q(w)$ should be negative.

Let $n\geq 1$, and let $w_1\in H^1(\R^2)$ be a radial function, with
$\|w_1\|_2=1$ and support in a ball of radius $1$.  Let $v\in
H^1(\R)$, with $\|v\|_2=1$, and choose as a trial function for $Q$
\beq
w(r,z)= (2c_n^2 \widetilde\mu_n)^{1/2} w_1( (2c_n^2
\widetilde\mu_n)^{1/2} r) \frac 1{L^{1/2}} v(z/L) \ ,
\eeq
with $c_n$ given in (\ref{cn}), and $L\geq 1$. 
Note that
$c_n=O(n^{-1/2})$ for large $n$ and, by (\ref{12}) and (\ref{En}),
\beq\label{mun}
\widetilde\mu_n\geq \inf_{\xx}\left\{
\frac{n^2}{r^2}+V(\x)\right\}\geq n \omega-C_\omega
\eeq
for any $\omega<\Omega_c$ and some constant $C_\omega$ depending
only on $\omega$. Therefore $c_n^2\widetilde \mu_n\geq \const$
independent of $n$ and $g$. Using (\ref{b1}) this implies that
\beq\label{a22}
\langle w| V |w\rangle \leq \const L^p \ .
\eeq
Denoting $T_r=\langle w_1|-\Delta_r|w_1\rangle$, $T_z=\langle
v|-\partial^2/\partial z^2|v\rangle$ and $M_n=\int d^2r r^{2n}
w_1$, we get the bound (note that the last term in (\ref{aaa}) is
zero for the $w$ under consideration, since $n\geq 1$)
\beq\label{qw}
Q(w)\leq n|\bfom|-\widetilde\mu_n+2c_n^2\widetilde\mu_n T_r + T_z
+\const L^p + 16\pi\frac g L \|r^{-2n}j_n(r)\|_\infty M_n \ .
\eeq
For some fixed $\eps>0$ and $n$ large enough, choose $L$ such that
the right side of (\ref{a22}) equals $\eps \widetilde\mu_n$. For
this choice of $L$,
\beq
\|r^{-2n} j_n\|_\infty\leq \widetilde C_\eps \frac L g
\widetilde\mu_n
\eeq
for some constant $\widetilde C_\eps$ depending only on $\eps$, as
can be seen from Lemmas~\ref{la1}--\ref{la3} and the fact that,
for large $\widetilde\mu_n$, $R\sim \widetilde\mu_n^{1/p}$ if
$\delta_n(R)=O(1)$ in (\ref{defdel}), again by our assumptions
(\ref{b1}) and (\ref{b2}) on $V$. Using these estimates in
(\ref{qw}) we obtain
\beq\label{qww}
Q(w)\leq n|\bfom|-\widetilde\mu_n\left( 1-\eps - 2c_n^2 T_r -
16\pi \widetilde C_\eps M_n\right)+ T_z \ .
\eeq
Since $|\bfom|<\Omega_c$ by assumption, and both $c_n^2$ and $M_n$
tend to zero as $n\to\infty$, we can choose $\eps$ appropriately
to conclude from (\ref{mun}) that the right side of (\ref{qww}) is
negative for $n$ large enough.

We have thus shown that for some $N_{\bfomm}$, independent of $g$, all
minimizers of the GP functional restricted to eigenfunctions of the
angular momentum $\bfom\cdot\bfL$ with eigenvalue $n\geq N_{\bfomm}$
are unstable and therefore cannot be absolute minimizers of the GP
functional.  Together with the result (\ref{aha}) above this proves
Theorem~\ref{T3}.
\end{proof}

\bigskip
\noindent {\it Acknowledgments.}
Helpful discussions with Elliott Lieb and Jakob Yngvason, as well as financial support by the Austrian Science Fund in the from of an Erwin
Schr\"odinger fellowship, are gratefully acknowledged.

\end{document}